\documentstyle[10pt]{article}

\def\bfl{\begin{flushleft}}
\def\efl{\end{flushleft}}
\def\bfr{\begin{flushright}}
\def\efr{\end{flushright}}
\def\bc{\begin{center}}
\def\ec{\end{center}}
\def\be{\begin{equation}}
\def\ee{\end{equation}}
\def\ba{\begin{eqnarray}}
\def\ea{\end{eqnarray}}
\def\nn{\nonumber }
\def\lb#1{\label{#1}}

\def\text#1{\mbox{#1}}
\def\drm{\text{d}}

\def\vphi{\vec\varphi}
\def\phik#1#2{\varphi_{#1}^{(s)\,#2}}

\def\Sign#1{\, \text{sign}\left(#1\right) }
\def\Der#1#2{\,\frac{\partial #1}{\partial #2}}

\def\Sech#1#2{\, \text{sch}^{#1}\left(#2 \right) }

\begin{document}

~\\

\bfr
Europhys. Lett. 49, No. 1, pp. 20-26 (2000)
~~~~~~~~\\
hep-th/9912064
~~~~~~~~\\
\efr
\bfl
{\LARGE \bf
Zero-brane approach to quantization of biscalar field\\
theory about topological kink-bell solution
}
\efl
\bc
~~\\
{\large Konstantin G. Zloshchastiev}\\
~~\\
E-mail: zlosh@email.com,
URL(s): http://zloshchastiev.webjump.com, http://zloshchastiev.cjb.net\\
~~\\

{\it
Received:  15 August 1999 (EPL), 8 December 1999 (LANL)}

\ec

~~\\

\abstract{\large
We study the properties of the topologically nontrivial
doublet solution
arisen in the biscalar theory with a fourth-power 
potential introducing an example of the spontaneous breaking
of symmetry.
We rule out the zero-brane (non-minimal point particle) action for this 
doublet as a particle with curvature.
When quantizing it as the theory with higher derivatives,
we calculate the quantum corrections to the 
mass of the doublet which could not be obtained by means
of the perturbation theory.
}

~\\
PACS number(s):  11.10.Lm, 11.15.Ex, 11.27.+d

\large

\section{Introduction}\lb{s-i}

The kink solution is known to be the soliton 
(more correctly, solitary wave) solution appearing 
in the 2D relativistic $\varphi^4$ one-scalar model and admitting an 
interpretation in terms of a particle.
Its classical and quantum properties are already studied in a number of
works \cite{dhn,kpp,col}.
However, the majority of recent theories contains the multiscalar
$\varphi^4$ theories (e.g., the scalar sector of the Weinberg-Salam model
contains such a complex biscalar theory).
Therefore, the increasing of internal space dimensions and 
studying of classical and 
quantum features of multiscalar theories seems to be of great necessity 
and interest.
Thus, the aim of this paper is to study the topologically nontrivial
solution of the biscalar $\varphi^4$ theory within the 
frameworks of the method developed (and applied to a one-scalar case) 
in Ref. \cite{kpp}. 
This approach consists in the 
constructing of brane action where the non-minimal
terms (first of all, depending on the world-volume curvature) are induced
by the field fluctuations in the neighborhood of the static solution.
Of course, these fluctuations are required to be reasonably small, 
and then the effective zero-brane action evidently arises 
after nonlinear reparametrization
of an initial theory when excluding zero field oscillations.

The paper is arranged as follows.
In Sec. \ref{s-kb} we study the kink-bell doublet solution and 
its properties on the classical level.
In Sec. \ref{s-ea} we perform the nonlinear parametrization of the 
initial action by means of the Bogolyubov transformation to the 
collective degrees of freedom.
After this, minimizing the action with respect to field fluctuations,
we remove zero modes and obtain the effective action
having the minimal and non-minimal (curvature) terms. 
Sec. \ref{s-q} is devoted to quantization of this action as a constrained
theory with higher derivatives.
We calculate the first excited level to obtain the
mass of the doublet with quantum corrections.

\section{Topological doublet}\lb{s-kb}

Let us consider the action describing $\varphi^4$ interaction of 
two spacetime scalar fields
\begin{equation}
S[\vphi] = \int  L(\vphi)\, \drm^2 x,              \lb{e1} 
\end{equation}
where 
\be                                                          \lb{e2}
L (\vphi) = \frac{1}{2} 
\sum_{a=1}^{2} (\partial_m \varphi_a) (\partial^m \varphi_a) -
U (\vphi),
\ee
and the potential part always can be rewritten in the following form:
\be                                                          \lb{e3}
U (\vphi) = \frac{1}{4} (\varphi_1^2 -1)^2 + 
\frac{m^2}{2} \varphi_2^2 + \frac{\lambda}{4} \varphi_2^4 +
\frac{1-2 m^2}{2\eta^2} \varphi_2^2 (\varphi_1^2-1).
\ee
The corresponding equations of motion,
\be                                                           \lb{e4}
\partial^m \partial_m \varphi_a + U_a(\vphi) = 0,
\ee
where
\[
U_a(\vphi) = \Der{U(\vphi)}{\varphi_a},~~
U_{ab}(\vphi) = \Der{^2\, U(\vphi)}{\varphi_a \partial \varphi_b},
\]
admit in the class of solitary waves,
\be                                                           \lb{e5}
\varphi_a(\rho) = \varphi_a
\left(
\frac{x-v t}{\sqrt{1-v^2}}
\right),
\ee
the following topologically nontrivial solution 
\be                                                          \lb{e6}
\phik{a}{} (\rho) =
\left\{
       \tanh{m \rho},\, \eta\, \text{sch}\, m \rho
\right\},
\ee
thereby
\[
\lambda \eta^4 = 1-2m^2(\eta^2 +1),
\]
and $\phik{a}{}$ are bound by the orbit equation
\be                                                          \lb{e7}
\phik{2}{2} + \eta^2 
\left(
\phik{1}{2} - 1
\right) = 0.
\ee
The first potential from the pair $\phik{a}{}$ is the 
well-known $\varphi^4$ kink 
whereas the second has a bell-like shape and seems to be topologically
trivial as such.
Nevertheless, the orbit equation provides strong stability of $\phik{2}{}$ 
by virtue of the topological nontriviality of $\phik{1}{}$ \cite{raj}.

The field doublet (\ref{e6}) has the localized energy density
\[                                                          
\varepsilon (x,t) = m^2
\Sech{4}{m \rho}
\left[
       1 + \eta^2 \sinh{\!^2 (m \rho)}
\right],                                                        
\]
and can be interpreted as the relativistic point particle with the
energy
\be                                                            \lb{e8}
E_{\text{class}} = \int\limits_{-\infty}^{+\infty}
\varepsilon (x,t)\ \drm x =
\frac{\mu}{\sqrt{1-v^2}},~~
\mu = \frac{2}{3} m \left( \eta^2 + 2 \right).
\ee
In subsequent sections we will deepen this interpretation both at
classical (considering field fluctuations) and quantum levels.

Finally, some comments upon relations between one- and two-scalar
$\varphi^4$ theories are in order.
The single scalar field theory with the coupling
constant of an arbitrary sign is described by the Lagrangian
\[
{\cal L}(\varphi) = 
\frac{1}{2} (\partial_m \varphi) (\partial^m \varphi)  - 
\frac{\lambda}{4} 
\biggl( \varphi^2 - \eta^2 \biggr)^2 + 
\frac{\lambda \eta^4}{8} 
( 1 - \xi ),                                     
\]
\[                                                           
\xi = \Sign{\lambda},~~~ \lambda,\,\eta \in \Re,
\]
where
the last term is introduced in such a way that 
the potential energy would be equal zero in the appropriate local minimum 
point.
At $\lambda < 0$ the state $\varphi=0$ 
is the most energetically favorable whereas at $\lambda > 0$
the states $\varphi=\pm \eta$ turn to be favorable thus realizing
the simplest spontaneous breaking of symmetry.
Further, at $\lambda>0$ this theory admits the kink as the 
only solution having localized energy.
However, at $\lambda<0$ the bell solution (like $\phik{2}{}$) appears
to be the only solution with localized energy.
With respect to energy the theory with $\lambda<0$ has both the unphysical
unbounded from below states and the physical ones 
(including the bell soliton) localized near a local minimum point 
which are stable against reasonably small field fluctuations.

Thus, in the one-scalar theory the kink and bell solutions cannot exist
together, whereas one can see that 
in the biscalar theory (\ref{e1}) they successfully coexist as 
a doublet within the same range of coupling parameters.

\section{Effective action}\lb{s-ea}

In this Section we will construct the nonlinear effective action of 
the biscalar $\varphi^4$ theory about the kink-bell doublet (\ref{e6}).
Let us change to the set of the collective coordinates 
$\{\sigma_0=s,\ \sigma_1=\rho\}$ such that
\be                                                      
x^m = x^m(s) + e^m_{(1)}(s) \rho,\ \              
\varphi_a(x,t) = \widetilde \varphi_a (\sigma),     
\ee
where $x^m(s)$ turn to be the coordinates of a (1+1)-dimensional point
particle, $e^m_{(1)}(s)$ is the unit spacelike vector orthogonal
to the world line.
Hence, the action (\ref{e1}) can be rewritten in new coordinates as
\be                                                         \lb{e10}
S[\widetilde \varphi] = 
\int L (\widetilde \varphi) \,\Delta \ \drm^2 \sigma,
\ee
where
\[
L (\widetilde \varphi) = \frac{1}{2} \sum_{a}
\left[
      \frac{(\partial_s \widetilde\varphi_a)^2}{\Delta^2} - 
                  (\partial_\rho \widetilde\varphi_a)^2
\right]
- U (\widetilde \varphi),
\]
\[
\Delta = \text{det} 
\left|
\left|
      \Der{x^m}{\sigma^k}
\right|
\right|
= \sqrt{\dot x^2} (1- \rho k),
\]
and $k$ is the curvature of a particle world line
\be                                                            \lb{e11}
k = \frac{\varepsilon_{mn} \dot x^m \ddot x^n}{(\sqrt{\dot x^2})^3},
\ee
where $\varepsilon_{m n}$ is the unit antisymmetric tensor.
This new action contains the two redundant degrees of freedom which 
eventually
lead to appearance of the so-called ``zero modes''.
To eliminate them we must constrain the model
by means of the condition of vanishing of the functional derivative with
respect to the doublet field fluctuations about some chosen static solution
(kink-bell in our case),
and in result we will obtain the required effective action \cite{kpp}.

So, the fluctuations of the fields $\widetilde\varphi_a (\sigma)$ in the 
neighborhood of the static solution $\phik{a}{} (\rho)$
are given by the expression
\be                                               
\widetilde\varphi_a (\sigma) = 
\phik{a}{} (\rho) + \delta \varphi_a (\sigma).
\ee
Substituting them into eq. (\ref{e10}) and considering the static
equation of motion (\ref{e4}) for $\phik{a}{}$ we have
\begin{eqnarray}                                          \lb{e13}
S[\delta \vphi] 
&=& \int d^2 \sigma \ 
   \Biggl\{\Delta 
        \Biggl[L(\vec\phik{}{}) +
               \frac{1}{2} \sum_{a}
               \Biggl( 
                  \frac{\left(\partial_s \ \delta \varphi_a \right)^2}
                       {\Delta^2} 
                  -  
                  \Bigl( 
                        \partial_{\rho}  \delta \varphi_a 
                  \Bigr)^2 - \nn\\
&&                  \sum_b U_{ab} (\vec\phik{}{}) 
                  \delta \varphi_a\, \delta \varphi_b
               \Biggr) 
        \Biggr]
        - k \sqrt{\dot x^2}\sum_a \phik{a}{\prime} \delta \varphi_a
        + O (\delta \varphi^3)                            
    \Biggr\} + \text{{surf. terms}},                                            
\ea
\[
L(\vec\phik{}{}) =  -
 \frac{1}{2} \sum_a \phik{a}{\prime\, 2} - U (\vec\phik{}{}),
\]
where prime means the derivative with respect to $\rho$.
Extremalizing this action with respect to 
$\delta \varphi_a$ one can obtain 
the system of equations in partial derivatives for field fluctuations:
\be
\left(
     \partial_s \Delta^{-1} \partial_s -
     \partial_{\rho} \Delta \partial_{\rho} 
\right) \delta\varphi_a
+\Delta \sum_{b} U_{ab} (\vec\phik{}{}) \delta\varphi_b
+ \phik{a}{\prime} k\sqrt{\dot{x}^2} =
O(\delta \varphi^2),
\ee
which has to be the constraint removing redundant degrees of
freedom.
Supposing $\delta\varphi_a (s,\rho) = k(s) f_a(\rho)$, in the 
linear approximations
$\rho k\ll 1$ (which naturally guarantees also
the smoothness of a world line at $\rho \to 0$) 
and $O(\delta\varphi^2)=0$ we obtain the system
of three ordinary derivative equations
\ba  
&&\frac{1}{\sqrt{\dot{x}^2}} \frac{d}{ds} 
\frac{1}{\sqrt{\dot{x}^2}} \frac{dk}{ds} +ck = 0,          \lb{e15}\\
&&-f''_a + \sum_b
\left( 
      U_{ab} (\vec\phik{}{}) - c \delta_{ab}
\right) f_b + \phik{a}{\prime} = 0.                        \lb{e16}
\ea
                                  
First of all, we are needed to find the solutions of (\ref{e16})
such that field fluctuations would be damping at both infinities 
(then the integral at a non-minimal term below will be finite).
In other words, we suppose the boundary conditions
\be                                                   \lb{e17}
f_a(+\infty) = f_a(-\infty) = O(1),
\ee 
which evidently correspond to the singular Stourm-Liouville problem 
describing bound states of some ``quantum'' system.
Further, varying the orbit equation (\ref{e7}) we obtain
the expression
\[
\frac{f_2}{f_1} 
= - \eta^2 \frac{\phik{1}{}}{\phik{2}{}}
= \frac{\phik{2}{\prime}}{\phik{1}{\prime}},
\]
one can separate $f_a$ in (\ref{e16}) by virtue of.
Thus, we have the two independent Stourm-Liouville problems which
can be resolved exactly and completely.
The following theorem has to be very useful in this connection.

{\it Theorem.}
Let us have the differential equation
\be                                               \lb{e18}
-f''(u) + 2\xi 
\left(
      3 X_0^2  - \frac{\xi+3}{4}
\right)f(u) - c f(u) + X_0' = 0,
\ee
where $c$ is an arbitrary constant, $\xi^2 = 1$,
\[
X_0 = \sqrt{\frac{1+\xi}{2} - \xi \text{sch}^2 u}.
\]
Then the corresponding Stourm-Liouville problem,
\be                                                   \lb{e19}
f(+\infty) = f(-\infty) = O(1),
\ee 
has the only solution for each $\xi$:
\[
f^{(\xi=+1)} = \frac{K \ \text{sinh} u + 1}{3 \text{cosh}^2 u}, \quad
f^{(\xi=-1)} = \frac{\text{sinh} u + B}{3 \text{cosh}^2 u},
\]
thereby the corresponding eigenvalues are
\[
c=3 \xi ,
\]
where $B$ and $K$ are arbitrary integration constants.\footnote{The 
proof is not very complicated but too bulky hence it is not presented 
here (see ref. \cite{zlo006} for the proof of similar theorem).}

With the use of this theorem we obtain that the solutions of 
system (\ref{e16}) provided (\ref{e17}) are the two sets
of eigenfunctions and eigenvalues which will be distinguished
by virtue of a hat index
\ba
&&f_{1\hat n} = \frac{m}{c_{\hat n}} \Sech{2}{m\rho}, \nn\\
&&f_{2\hat n} =- \frac{\eta m}{c_{\hat n}} 
\frac{\tanh{m\rho}}{\cosh{m\rho}}, \\
&&c_{\hat n} = 3 (-1)^n m^2,~~n=0,1. \nn
\ea
Substituting the found functions back in the action (\ref{e13}),
we can rewrite it in the explicit p-brane form
\be                                \lb{e28}
S_{\text{eff}} = 
S_{\text{eff}}^{\text{(class)}} + S_{\text{eff}}^{\text{(fluct)}} =
- \int \drm s \sqrt{\dot x^2} 
\left(
       \mu + \alpha k^2
\right),
\ee
where
\[
\mu = - \int\limits_{-\infty}^{+\infty} \drm \rho (1 - \rho k)
L (\vec\phik{}{}) =
\int\limits_{-\infty}^{+\infty} \varepsilon (x,t)\  \drm \rho,
\]
see eq. (\ref{e8}), and
\be
\alpha = \frac{1}{2} \sum_a
\int\limits_{-\infty}^{+\infty} 
f_{a \hat n} \phik{a}{\prime} \ \drm \rho
=\frac{\mu}{2 c_{\hat n}} = (-1)^n \frac{\eta^2 +2}{9 m}.
\ee
The action (\ref{e28}) yields the equations of motion for the 
kink-bell doublet field solution as a particle with curvature
\be
\frac{1}{\sqrt{\dot x^2}}
\frac{\drm}{\drm s}
\frac{1}{\sqrt{\dot x^2}}
\frac{\drm k}{\drm s} +
      \frac{\mu - \alpha k^2}{2\alpha} k = 0,
\ee
hence one can see that eq. (\ref{e15}) was nothing
but this equation in the linear approximation $k \ll 1$,
as was expected.

Thus, the considering of field fluctuations naturally 
leads to the splitting of the kink-bell 
doublet into the two subtypes in dependence on the constant
before the non-minimal term $k^2$.
However, when studying the quantum aspects of the model we will find that
the absolute value of mass (even with quantum corrections) does not 
depend on the sign of $c$, 
the energy of the non-minimal particle with $n=1$ lies in the
lower energy continuum and hence can be interpreted in terms of 
antiparticles.
Therefore, below we will assume $n=0$, i.e., $\alpha = \mu/6 m^2$.

\section{Quantization}\lb{s-q}

In the previous section we obtained a classical 
effective action for the model in question.
Thus, to quantize it we must consecutively construct the 
Hamiltonian structure of dynamics of the point particle with curvature.
From eqs. (\ref{e11}) and (\ref{e28}) one can see that we have the 
theory with higher derivatives \cite{dhot,ply}.
Hence, below we will treat the coordinates and momenta as the 
canonically independent coordinates of phase space.
Besides, the Hessian matrix, constructed 
from the derivatives with respect to accelerations,
appears to be singular that says about the presence of the
constraints on phase variables of the theory.
Following the quantization procedure proposed in ref. \cite{kpp} and
developed in ref. \cite{zlo006} one can obtain the equation for 
discrete mass spectrum in the form
\be
\varepsilon =  \sqrt{B(B-1)} (n + 1/2) 
+ O (\hbar^2),\ \ n=0,\ 1,\ 2, ...,
\ee
where
\[
B= \frac{8\sqrt{2}}{3}
       \sqrt{
            \frac{\eta^2 + 2}{m} M  
            }, \; \;                                             
\varepsilon = \frac{8}{3} \frac{\mu^2}{m^2} 
\left(
      1 - \frac{M}{\mu}
\right), 
\]
and $M$ is the total mass of the kink-bell doublet as a 
non-minimal point particle with curvature.

Further, again following \cite{zlo006},
the first quantum correction to particle masses will be
determined by the lower energy of oscillations:
\be                                                          \lb{e51}
\varepsilon = \frac{1}{2} \sqrt{B(B-1)} + O (\hbar^2),
\ee
that gives the algebraic equation for $M$ as a function of $m$
and $\mu$.

We can easily resolve it in the approximation of weak coupling.
Assuming $\lambda \sim m^2/\eta^2 \to 0$ (or, equivalently, 
$B \gg 1$) in such a way that the value $m^2$ remains to be nonzero
we find eq. (\ref{e51}) in the form
\be                                                         
\varepsilon = \frac{B}{2} + O (\lambda\hbar^2),
\ee
which after all yields
\be
(M-\mu)^2 = \frac{3}{4} \frac{m^2}{\mu} M + O (\lambda\hbar^2).
\ee
Then one can seek for mass in the form $M=\mu+\delta$ 
($\delta \ll \mu$), and 
finally we obtain
\be
M = \mu \pm \frac{\sqrt{3}}{2} m + O(\lambda\hbar^2).
\ee
Thus, the mass of the doublet boson with quantum corrections reads
in first approximation
\be
M = \frac{4\eta^2 + 8 \pm 3\sqrt{3}}{6} m,
\ee
hence one can see that the main term in this expression turns to
be singular at $\lambda \to 0$, therefore, the obtained results are 
nonperturbative and can not be ruled out 
through the $\lambda$-series of the perturbation theory.

\def\CMPh{Commun. Math. Phys.}
\def\JMP{J. Math. Phys.}
\def\JPh{J. Phys.}
\def\CJP{Czech. J. Phys.}
\def\FP{Fortschr. Phys.}
\def\LMPh {Lett. Math. Phys.}
\def\MPL {Mod. Phys. Lett.}
\def\NPh  {Nucl. Phys.}
\def\PhE  {Phys.Essays}
\def\PhL  {Phys. Lett.}
\def\PhR  {Phys. Rev.}
\def\PhRL {Phys. Rev. Lett.}
\def\PhRp {Phys. Rep.}
\def\NCim {Nuovo Cimento}
\def\NuPB {Nucl. Phys.}
\def\TMF {Teor. Mat. Fiz.}
\def\GRG {Gen. Relativ. Gravit.}
\def\CQG {Class. Quantum Grav.}
\def\prp {report}
\def\Prp {Report}

\def\jn#1#2#3#4#5{{#1}{#2} {#3} {(#5)} {#4}}   

\def\boo#1#2#3#4#5{ #1 ({#2}, {#3}, {#4}){#5}}  

\def\prpr#1#2#3#4#5{{``#1,''} {#2}{ #3}{#4}, {#5}}                

~\\
~\\
~\\
Regrettably, ref. \cite{ply} was added when the paper had already 
been published in EPL.

\end{document}